\documentclass[twoside]{article}
\catcode`\@=11
\long\def\@makefntext#1{
\protect\noindent \hbox to 3.2pt {\hskip-.9pt
$^{{\eightrm\@thefnmark}}$\hfil}#1\hfill}               %CAN BE USED

\def\@makefnmark{\hbox to 0pt{$^{\@thefnmark}$\hss}}    %ORIGINAL

\def\ps@myheadings{\let\@mkboth\@gobbletwo
\def\@oddhead{\hbox{}
\rightmark\hfil\eightrm\thepage}
\def\@oddfoot{}\def\@evenhead{\eightrm\thepage\hfil
\leftmark\hbox{}}\def\@evenfoot{}
\def\sectionmark##1{}\def\subsectionmark##1{}}

\oddsidemargin=\evensidemargin
\usepackage{graphics}
\newcounter{sectionc}\newcounter{subsectionc}\newcounter{subsubsectionc}
\renewcommand{\section}[1] {\vspace{12pt}\addtocounter{sectionc}{1}
\setcounter{subsectionc}{0}\setcounter{subsubsectionc}{0}\noindent
        {\tenbf\thesectionc. #1}\par\vspace{5pt}}
\renewcommand{\subsection}[1] {\vspace{12pt}\addtocounter{subsectionc}{1}
        \setcounter{subsubsectionc}{0}\noindent
        {\bf\thesectionc.\thesubsectionc. {\kern1pt \bfit #1}}\par\vspace{5pt}}
\renewcommand{\subsubsection}[1] {\vspace{12pt}\addtocounter{subsubsectionc}{1}
        \noindent{\tenrm\thesectionc.\thesubsectionc.\thesubsubsectionc.
        {\kern1pt \tenit #1}}\par\vspace{5pt}}
\newcommand{\nonumsection}[1] {\vspace{12pt}\noindent{\tenbf #1}
        \par\vspace{5pt}}

\newcounter{appendixc}
\newcounter{subappendixc}[appendixc]
\newcounter{subsubappendixc}[subappendixc]
\renewcommand{\thesubappendixc}{\Alph{appendixc}.\arabic{subappendixc}}
\renewcommand{\thesubsubappendixc}
        {\Alph{appendixc}.\arabic{subappendixc}.\arabic{subsubappendixc}}

\renewcommand{\appendix}[1] {\vspace{12pt}
        \refstepcounter{appendixc}
        \setcounter{figure}{0}
        \setcounter{table}{0}
        \setcounter{lemma}{0}
        \setcounter{theorem}{0}
        \setcounter{corollary}{0}
        \setcounter{definition}{0}
        \setcounter{equation}{0}
        \renewcommand{\thefigure}{\Alph{appendixc}.\arabic{figure}}
        \renewcommand{\thetable}{\Alph{appendixc}.\arabic{table}}
        \renewcommand{\theappendixc}{\Alph{appendixc}}
        \renewcommand{\thelemma}{\Alph{appendixc}.\arabic{lemma}}
        \renewcommand{\thetheorem}{\Alph{appendixc}.\arabic{theorem}}
        \renewcommand{\thedefinition}{\Alph{appendixc}.\arabic{definition}}
        \renewcommand{\thecorollary}{\Alph{appendixc}.\arabic{corollary}}
        \renewcommand{\theequation}{\Alph{appendixc}.\arabic{equation}}
        \noindent{\tenbf Appendix \theappendixc #1}\par\vspace{5pt}}
\newcommand{\subappendix}[1] {\vspace{12pt}
        \refstepcounter{subappendixc}
        \noindent{\bf Appendix \thesubappendixc. {\kern1pt \bfit #1}}
        \par\vspace{5pt}}
\newcommand{\subsubappendix}[1] {\vspace{12pt}
        \refstepcounter{subsubappendixc}
        \noindent{\rm Appendix \thesubsubappendixc. {\kern1pt \tenit #1}}
        \par\vspace{5pt}}

\newcommand{\textlineskip}{\baselineskip=13pt}
\newcommand{\smalllineskip}{\baselineskip=10pt}

\def\eightcirc{
\begin{picture}(0,0)
\put(4.4,1.8){\circle{6.5}}
\end{picture}}
\def\eightcopyright{\eightcirc\kern2.7pt\hbox{\eightrm c}}

\newcommand{\copyrightheading}[1]
        {\vspace*{-2.5cm}\smalllineskip{\flushleft
        {\footnotesize International Journal of Modern Physics B: {\bf Proceedings  of 
the Second International  Conference on New Theories, Discoveries and Applications of
Superconductors and Related Materials, Las Vegas, May 31-June 1 (1999)}
  {} #1}\\
        {\footnotesize $\eightcopyright$\, World Scientific Publishing
         Company}\\
         }}

\newcommand{\publisher}[2]{{\begin{center}\footnotesize\smalllineskip
        Received #1\\
        Revised #2
        \end{center}
        }}

\def\abstracts#1#2#3{{
        \centering{\begin{minipage}{4.5in}\baselineskip=10pt\footnotesize
        \parindent=0pt #1\par
        \parindent=15pt #2\par
        \parindent=15pt #3
        \end{minipage}}\par}}

\renewenvironment{thebibliography}[1]                   %ALL CHANGES DD 13/3/92
        {\frenchspacing
         \ninerm\baselineskip=11pt
         \begin{list}{\arabic{enumi}.}
        {\usecounter{enumi}\setlength{\parsep}{0pt}
         \setlength{\leftmargin 12.7pt}{\rightmargin 0pt} %FOR 1--9 ITEMS
         \setlength{\itemsep}{0pt} \settowidth
        {\labelwidth}{#1.}\sloppy}}{\end{list}}

\newcounter{itemlistc}
\newcounter{romanlistc}
\newcounter{alphlistc}
\newcounter{arabiclistc}

\newcommand{\fcaption}[1]{
        \refstepcounter{figure}
        \setbox\@tempboxa = \hbox{\footnotesize Fig.~\thefigure. #1}
        \ifdim \wd\@tempboxa > 5in
           {\begin{center}
        \parbox{5in}{\footnotesize\smalllineskip Fig.~\thefigure. #1}
            \end{center}}
        \else
             {\begin{center}
             {\footnotesize Fig.~\thefigure. #1}
              \end{center}}
        \fi}

\newcommand{\tcaption}[1]{
        \refstepcounter{table}
        \setbox\@tempboxa = \hbox{\footnotesize Table~\thetable. #1}
        \ifdim \wd\@tempboxa > 5in
           {\begin{center}
        \parbox{5in}{\footnotesize\smalllineskip Table~\thetable. #1}
            \end{center}}
        \else
             {\begin{center}
             {\footnotesize Table~\thetable. #1}
              \end{center}}
        \fi}

\def\@citex[#1]#2{\if@filesw\immediate\write\@auxout
        {\string\citation{#2}}\fi
\def\@citea{}\@cite{\@for\@citeb:=#2\do
        {\@citea\def\@citea{,}\@ifundefined
        {b@\@citeb}{{\bf ?}\@warning
        {Citation `\@citeb' on page \thepage \space undefined}}
        {\csname b@\@citeb\endcsname}}}{#1}}

\newif\if@cghi
\def\cite{\@cghitrue\@ifnextchar [{\@tempswatrue
        \@citex}{\@tempswafalse\@citex[]}}
\def\citelow{\@cghifalse\@ifnextchar [{\@tempswatrue
        \@citex}{\@tempswafalse\@citex[]}}
\def\@cite#1#2{{$\null^{#1}$\if@tempswa\typeout
        {IJCGA warning: optional citation argument
        ignored: `#2'} \fi}}

\def\pmb#1{\setbox0=\hbox{#1}
        \kern-.025em\copy0\kern-\wd0
        \kern.05em\copy0\kern-\wd0
        \kern-.025em\raise.0433em\box0}

\def\fnt#1#2{\footnotetext{\kern-.3em
        {$^{\mbox{\scriptsize #1}}$}{#2}}}

\def\fpage#1{\begingroup
\voffset=.3in
\thispagestyle{empty}\begin{table}[b]\centerline{\footnotesize #1}
        \end{table}\endgroup}

\def\runninghead#1#2{\pagestyle{myheadings}
\markboth{{\protect\footnotesize\it{\quad #1}}\hfill}
{\hfill{\protect\footnotesize\it{#2\quad}}}}
\font\tenrm=cmr10
\font\tenit=cmti10
\font\tenbf=cmbx10
\font\bfit=cmbxti10 at 10pt
\font\ninerm=cmr9
\font\nineit=cmti9
\font\ninebf=cmbx9
\font\eightrm=cmr8

\def\qed{\hbox{${\vcenter{\vbox{                        %HOLLOW SQUARE
   \hrule height 0.4pt\hbox{\vrule width 0.4pt height 6pt
   \kern5pt\vrule width 0.4pt}\hrule height 0.4pt}}}$}}

\def\bsc{{\sc a\kern-6.4pt\sc a\kern-6.4pt\sc a}}       %LATEX LOGO
\def\bflatex{\bf L\kern-.30em\raise.3ex\hbox{\bsc}\kern-.14em
T\kern-.1667em\lower.7ex\hbox{E}\kern-.125em X}

\begin{document}

\runninghead{Relationship between Conductivity and Phase Coherence Length in
Cuprates}
{Relationship between Conductivity and Phase Coherence Length in Cuprates}

\normalsize\textlineskip
\thispagestyle{empty}
\setcounter{page}{1}

\copyrightheading{}                     %{Vol. 0, No. 0 (1993) 000---000}

\vspace*{0.88truein}

\fpage{1}
\centerline{\bf Relationship between Conductivity and}
\vspace*{0.035truein}
\centerline{\bf  Phase Coherence Length in Cuprates}
%\vspace*{0.035truein}
%\centerline{\bf }
\vspace*{0.37truein}
\centerline{\footnotesize C. C. ALMASAN, G. A. LEVIN, E. CIMPOIASU,  T.
STEIN, and C. L. ZHANG}
\vspace*{0.015truein}
\centerline{\footnotesize\it Department of Physics, Kent State University}
\baselineskip=10pt
\centerline{\footnotesize\it Kent, Ohio 44242, USA}
\vspace*{10pt}
\centerline{\footnotesize M. C. DeANDRADE  and M. B. MAPLE}
\vspace*{0.015truein}
\centerline{\footnotesize\it Department of Physics and Institute for Pure
and Applied Physical Sciences}
\baselineskip=10pt
\centerline{\footnotesize\it University of California, San Diego}
\baselineskip=10pt
\centerline{\footnotesize\it La Jolla, California, 92093, USA}
\vspace*{10pt}
\centerline{\footnotesize HONG ZHENG, A. P. PAULIKAS, and B. W. VEAL}
\vspace*{0.015truein}
\centerline{\footnotesize\it Materials Science Division, Argonne National
Laboratory}
\centerline{\footnotesize\it Argonne, Illinois 60439, USA}
\baselineskip=10pt
\vspace*{0.225truein}
\publisher{31 May 1999}{(revised date)}

\vspace*{0.21truein}
\abstracts{The large ($10^2 - 10^5$) and strongly temperature dependent resistive 
anisotropy $\eta =
(\sigma_{ab}/\sigma_c)^{1/2}$ of cuprates perhaps  holds the
key to understanding  their normal state in-plane $\sigma_{ab}$ and out-of-plane
$\sigma_{c}$ conductivities.   It can be shown that $\eta$ is
determined by the ratio of the phase coherence lengths $\ell_i$ in the respective directions:
$\sigma_{ab}/\sigma_c =
\ell_{ab}^2/\ell_{c}^2$.  In layered crystals in which the out-of-plane transport is  incoherent,
$\ell_{c}$ is fixed, equal to the interlayer spacing.
As a result, the T-dependence of $\eta$ is determined by  that of
$\ell_{ab}$, and vice versa,  the in-plane phase coherence length  can be obtained directly by
measuring  the resistive anisotropy.
We present data for hole-doped $YBa_2Cu_3O_y$ ($6.3
< y < 6.9$) and $Y_{1-x}Pr_xBa_2Cu_3O_{7-\delta }$ ($0 < x \leq 0.55$)
and show that $\sigma_{ab}$ of  crystals with different  doping levels
can be  well described by a two parameter universal function of
the in-plane phase coherence length.
In  the electron-doped $Nd_{2-x}Ce_{x}CuO_{4-y}$, the dependence $\sigma_{ab}(\eta )$
indicates a crossover from incoherent to coherent transport in the c-direction.}{}{}
%\vspace*{10pt}
%\keywords{The contents of the keywords}

\textlineskip                  %) USE THIS MEASUREMENT WHEN THERE IS
\vspace*{12pt}                 %) NO SECTION HEADING

%\vspace*{1pt}\textlineskip      %) USE THIS MEASUREMENT WHEN THERE IS
%\section{General Appearance}    %) A SECTION HEADING
\vspace*{-0.5pt}
\noindent

We present data on  the normal state in-plane
$\sigma_{ab}$ and out-of-plane
$\sigma_{c}$ conductivities of  anisotropic layered crystals as diverse as
hole-doped
$YBa_{2}Cu_{3}O_{y}$ ($6.35<y<6.93$) and $Y_{1-x}Pr_xBa_2Cu_3O_{7-\delta }$
($0 < x \leq 0.55$), and electron-doped
$Nd_{2-x}Ce_{x}CuO_{4-y}$.
The conductivities of oxygen deficient $YBa_2Cu_3O_{y}$ single
crystals  were measured  using the four-point method as
well as a multiterminal technique\cite{Jiang}, in zero field and in
a field of $14\;T$
in order to suppress superconductivity and reveal the normal state down to lower
temperatures. The conductivities of $Y_{1-x}Pr_xBa_2Cu_3O_{7-\delta }$ and $Nd_{2-x}Ce_{x}CuO_{4-y}$  single
crystals were measured using  the multiterminal 
method.

Recently, it has been shown \cite{PRL,MOS99} that the ratio of the conductivities is given by the
ratio of the phase coherence lengths,  hereafter called Thouless length (TL), in the respective
directions: 
\begin{equation}
\frac{\sigma_{ab}}{\sigma_{c}}=\frac{\ell^2}{\ell_c^2}.
\end{equation}
Here  $\ell$ is the in-plane TL ( subscript $"ab"$ is omitted)  and, for simplicity, we  assume
isotropic in-plane conduction,
$\sigma_{a}=\sigma_{b}\equiv \sigma_{ab}$.  The incoherent out-of-plane transport means that
the c-axis TL  ($\ell_c$) is temperature independent, equal to the interlayer spacing $\ell_0$.
Consequently, the in-plane TL   is given directly by the measured anisotropy; i.e., 
$\ell=(\rho_c/\rho_{ab})^{1/2}\ell_0$. Therefore, $\sigma (\ell)$ can be
obtained by plotting $\sigma_{ab}$ vs  $(\rho_c/\rho_{ab})^{1/2}\ell_0$.

Figures 1(a) and 1(b) are plots of $\sigma_{ab}$ vs  $
(\rho_c/\rho_{ab})^{1/2}\ell_0$ ($\ell_0=11.7\AA$) of $YBa_{2}Cu_{3}O_{y}$
and
$Y_{1-x}Pr_xBa_2Cu_3O_{7-\delta }$, respectively. The open  symbols represent raw data
corresponding to different doping. The dashed curves are ``isotherms" joining points
corresponding to the same temperature.  All data were taken within the T-range $T_c<T\le 300\;K$.
Notice that, at a given temperature,
the anisotropy  changes nonmonotonically as a function of doping, reaching a
maximum around $ y\approx  6.42$ in $YBa_{2}Cu_{3}O_{y}$ and
$x\approx 0.42$ in $Y_{1-x}Pr_xBa_2Cu_3O_{7-\delta }$.
According to Eq. (1), the anisotropy is determined by the in-plane TL
which, in turn,  correlates    with the localization length on the
insulating side and the correlation length on the metallic side  of  the metal-insulator
transition (MIT).\cite{PRL}
Both of these  length scales   increase   on approaching  the MIT
and so does the anisotropy, indicating  that the MIT takes place close to  $y= 6.42$
in $YBa_{2}Cu_{3}O_{y}$ and  $x=0.42$ in $Y_{1-x}Pr_xBa_2Cu_3O_{7-\delta}$.

An idea  outlined in Ref. 2
is that  variation with doping of the number of carriers and of the amount
of disorder  do not
alter fundamentally the
$\sigma (\ell )$ dependence. The effect  of these changes can be
incorporated into
two constants, one  of which, $\bar\sigma $,  normalizes the magnitude of
the in-plane conductivity,
and the other, $\bar\ell $,   normalizes the in-plane Thouless length:
\begin{eqnarray}
\frac{\sigma_{ab}}{\bar\sigma }=f\left (\frac{\ell}{\bar\ell}\right
);\;\;\;f(1)=1.
\end{eqnarray}
Here $f(y)$ is the same function
for a given class of single crystals, such as $YBa_{2}Cu_{3}O_{y}$.
Varying dopant concentration  only changes the values of the
normalization constants
$\bar\sigma$ and $ \bar\ell$, shifting the respective segments  of
$\sigma_{ab}(\ell )$
along a  common trajectory.

The continuous curves indicated in Figs. 1(a) and 1(b) by the filled
symbols were generated by shifting the segments corresponding to different oxygen and Pr
concentrations, respectively,  parallel to themselves (as indicated
by the arrows),   matching the values and the slopes of the {\it overlapping}
segments. The data for  the $y=6.93$  and $x=0.53$ samples  were   used as references in 
Figs. 1(a) and 1(b),
respectively. On a log-log plot, such  shifts are equivalent to  a change of  the normalization
constants $\bar\sigma$ and $\bar\ell$.    Through this
procedure, we obtained the  trajectory
$\sigma_{ab}(\ell )$ given by Eq. (2) with the values  of
$\bar\sigma $ and $\bar\ell $ corresponding to  the reference samples. 
The branching point  in Fig. 1(a)
reflects the MIT.  The branching point  is absent in Fig. 1(b) due to the lack
of data for   samples close to the MIT, on its ``metallic side".

\begin{figure}[htbp]
\vspace*{-0.14in}
%\vspace*{-13pt}
%\centerline{\vbox{\hrule width 5cm height0.001pt}}
%\vspace*{1.5truein}             %ORIGINAL SIZE=1.6TRUEIN x 100% - 0.2TRUEIN
%\vskip 0.1in
%\hspace*{0.95in}
\hspace*{-0.15in}
%\special{pictfile Fig1.pict scaled 0.54}
\includegraphics*{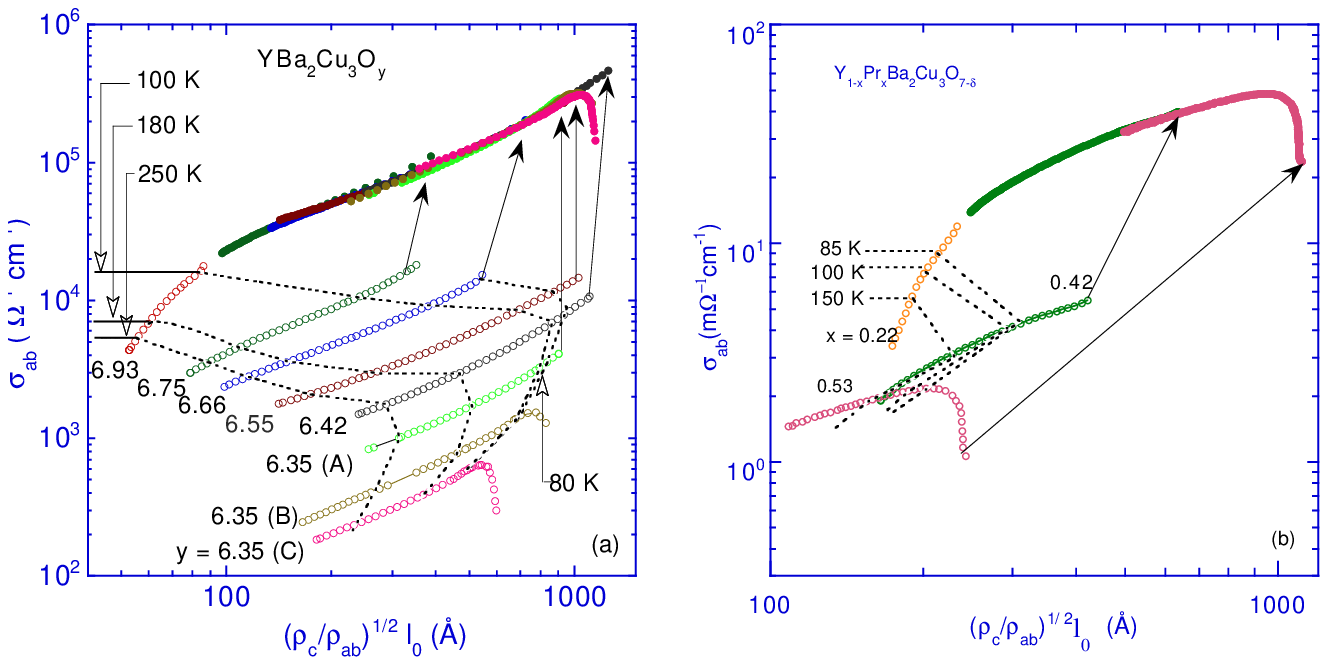}
%\centerline{\vbox{\hrule width 5cmheight0.001pt}}
\vspace*{13pt}
\fcaption{In-plane conductivity $\sigma_{ab}$  of (a) $YBa_2Cu_3O_{y}$ and (b) 
$Y_{1-x}Pr_xBa_2Cu_3O_{7-\delta}$
single crystals plotted vs phase coherence
length defined as $\ell= (\rho_c/\rho_{ab})^{1/2}\ell_0$, with
$\ell_0=11.7\AA$.
The open symbols correspond to raw data. The filled symbols which form a
trajectory [Eq. (2)] are obtained by shifting the open symbol
segments parallel to themselves
as indicated by arrows.  The dashed lines are ``isotherms",  joining points
corresponding to the same temperature. For all samples, anisotropy monotonically increases
with decreasing temperature. 
}
\end{figure}

The fact that the hole-doped cuprates follow so well the scaling relationship (2)
indicates that, indeed, the out-of-plane TL is T-independent in these crystals. However, there  
must  also be systems  for which a crossover from incoherent to coherent out-of-plane
transport takes place.   In such an intermediate regime, the TL in the c-direction becomes
T-dependent, but is  much smaller and  increases  at a smaller rate with decreasing temperature
than the in-plane TL. As a  result,  the anisotropy 
$\sigma_{ab}/\sigma_{c}=\ell_{ab}^2/\ell_{c}^2$  would still increase  with decreasing $T$
over an extended T-range.  
We looked for an example of such a regime in the 
data of electron-doped  $Nd_{2-x}Ce_{x}CuO_{4-y}$.  
These crystals have very large anisotropy ($\sigma_{ab}/\sigma_{c}\sim 10^4 -10^5$),
similar to the most anisotropic hole-doped cuprates, however, both
 $\rho_{ab}(T)$ and $\rho_c(T)$
are metallic at all temperatures $T>T_c$. The
T-dependence of $\rho_{ab}$
is quadratic, similar to conventional 3D metals described by
Boltzmann-Landau transport theory.  We  studied several single crystals with two nominal
Ce concentrations of $x=0.08$ and $0.29$, which correspond to underdoped and
overdoped regime, respectively. Within each group, the resistivities varied 
due to uncontrolled variations in the oxygen content.  

When the out-of-plane TL becomes  temperature dependent, $\eta\equiv (\rho_c/\rho_{ab})^{1/2}$ 
no longer reflects 
the $T$-dependence of in-plane TL, as in the case of hole-doped systems. 
Let us assume that both TLs are determined by diffusion, so that
\begin{equation}
\ell_{ab}^2= D_{\|}\tau_{\varphi}+\xi^2;\;\; \ell_{c}^2= D_{\perp}\tau_{\varphi}+\ell_0^2.
\end{equation}

Here $D_{\|}$ and $D_{\perp}$ are the components of the diffusion tensor, 
$\tau_{\varphi}(T)$ is the decoherence time, and $\xi$ and $\ell_0$ are cutoffs.
These  empirical cutoffs  ensure that, in the {\it high temperature} limit
[the limit of small $\tau_{\varphi}(T)$], $\ell_{ab}$ does not scale to zero with decreasing
$\tau_{\varphi}$ and $\ell_c$ saturates at the value of the interlayer
distance $\ell_0$ (Refs. 2, 3). The anisotropy is then given by
\begin{equation}
\frac{\sigma_{ab}}{\sigma_{c}}=\frac{D_{\|}\tau_{\varphi}+\xi^2}{D_{\perp}\tau_{\varphi}+\ell_0^2}.
\end{equation}
\begin{figure}[htbp]
\vspace*{-0.14in}
%\vspace*{13pt}
%\centerline{\vbox{\hrule width 5cm height0.001pt}}
%\vspace*{1.0truein}             %ORIGINAL SIZE=1.6TRUEIN x 100% - 0.2TRUEIN
%\vskip 1.8in
\hspace*{-0.35in}
%\special{pictfile Fig2.pict scaled 0.56}
\includegraphics*{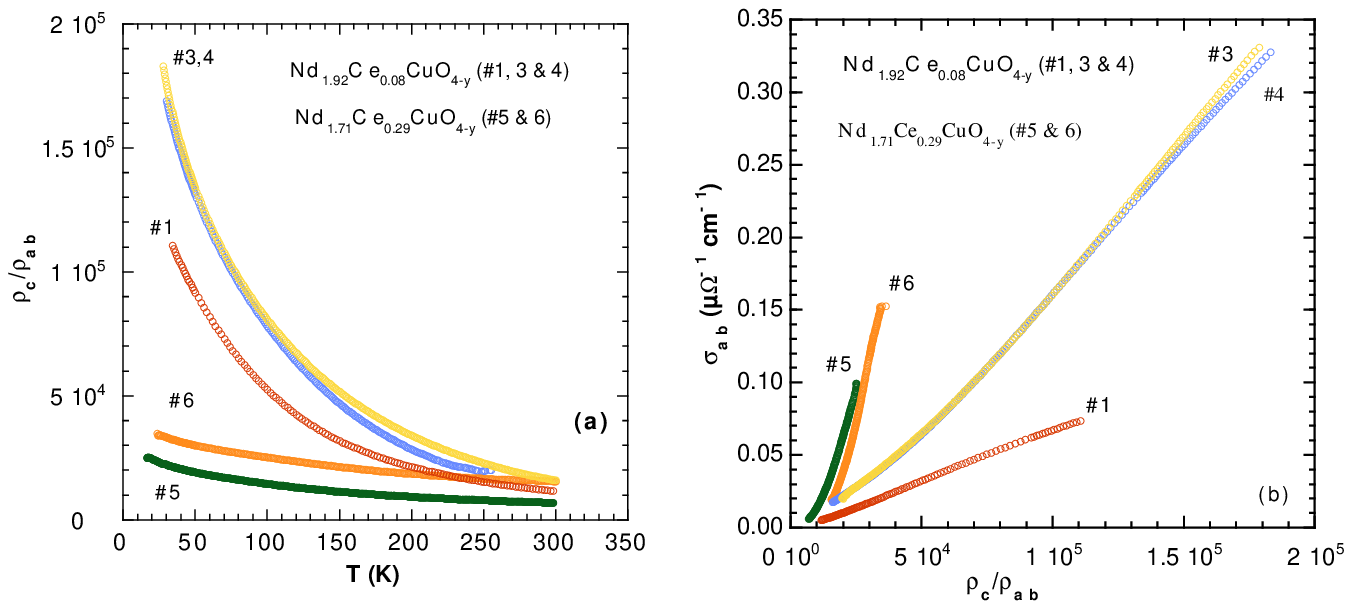}
%\centerline{\vbox{\hrule width 5cm height0.001pt}}
\vspace*{-7.5in}
\fcaption{ (a) Anisotropy $\rho_c/\rho_{ab}$ of $Nd_{2-x}Ce_{x}CuO_{4-y}$ single crystals vs
temperature $T$.  (b) In-plane conductivity $\sigma_{ab}$  plotted vs $\eta^2\equiv
\rho_c/\rho_{ab}$.   In both figures samples  $\#1-\;4$ are nominally underdoped while
$\#5$ and $6$ are nominally overdoped. Most  curves exhibit a more rapid than quadratic
increase  of $\sigma_{ab}$ with
$\eta$,  which is a signature of the crossover from incoherent to coherent
interlayer transport.
}
\end{figure}

The fact that the anisotropy in $Nd_{2-x}Ce_{x}CuO_{4-y}$  
increases
with decreasing temperature [Fig. 2(a)]  indicates that $D_{\|}\gg D_{\perp}$ 
and $D_{\perp}\tau_{\varphi}\sim \ell_0^2$. In the limit $T\rightarrow 0$, when
$D_{\perp}\tau_{\varphi}\gg \ell_0^2$, the anisotropy should saturate at the level $D_{\|}/
D_{\perp}$.

The Drude-type conductivity is given by  $\sigma_{ab}= k\tau$ ($k=const$), where $\tau$ is the
relaxation time of the distribution function. Assuming that $\tau$ and $\tau_{\varphi}$ are proportional
to each other [$\tau_{\varphi}=A\tau $ ($A=const $)], one gets from Eq. (4) the following expression
for
$\sigma_{ab}(\eta )$:
\begin{equation}
\sigma_{ab}=\sigma_{0}\frac{\eta^2-(\xi^2/\ell_0^2)}{1-\epsilon \eta^2},
\end{equation}
where $\epsilon =D_{\perp}/D_{\|}\ll 1$, and $\sigma_{0}=k\ell_0^2/D_{\|}A$.
Equation (5) shows that the signature of crossover to coherent out-of-plane transport is a
more rapid than {\it quadratic} increase of $\sigma_{ab}$ with $\eta$.
Note that the incoherent regime ($\ell_c =\ell_0$) would lead to 
$\sigma_{ab}\propto \eta^2$ for difusive motion  and to 
linear $\sigma_{ab}(\eta )\propto \eta -(\xi /\ell_0)$ for ballistic motion of the electrons (which is
the case in optimally doped $YBa_{2}Cu_{3}O_{y}$) (see Ref. 3). 
The divergence of $\sigma_{ab}(\eta )$ when
$\eta^2\rightarrow 1/\epsilon$ corresponds to the saturation of the anisotropy at low temperatures 
at the level $D_{\|}/ D_{\perp}$
and, correspondingly, to  the fully developed coherent transport in all directions. Our analysis,
therefore,  is   especially useful to identify the intermediate regime of out-of-plane coherence when
the anisotropy is still strongly T-dependent as is the case of $Nd_{2-x}Ce_{x}CuO_{4-y}$.

All  curves (except perhaps  sample $\#1$) in Fig. 2(b) display stronger than quadratic
$\sigma_{ab}(\eta )$ dependence indicating that, indeed, in these samples there is a crossover
to coherent out-of-plane transport.
Note that the overdoped crystals ($\# 5$ and $6$) have  appreciably lower 
anisotropy and slower rate of  its increase with decreasing temperature than
the  underdoped crystals ($\# 1-4$) [Fig. 2(a)]. 
This reflects that the ratio $ \epsilon
=D_{\perp}/D_{\|}$ increases  with overdoping. Correspondingly, the rate of
increase of $\sigma_{ab}(\eta )$ is much higher  in overdoped than in underdoped crystals [Fig.
2(b) and Eq. (5)]. Therefore, even though the underdoped
$Nd_{2-x}Ce_{x}CuO_{4-y}$ crystals are already in the intermediate regime of coherence
(unlike the hole doped crystals), the development of interlayer coherence accelerates visibly with
overdoping. However, by fitting the data to Eq. (5) we estimate that even in the overdoped samples,
the coherence barely extends from a given $CuO_2$ layer to its next-nearest neighbor even at
temperature as low as
$50\;K$.
 
In summary, we have shown that  both  $YBa_{2}Cu_{3}O_{y}$ and
$Y_{1-x}Pr_xBa_2Cu_3O_{7-\delta }$  systems are characterized by a universal
functional dependence of the in-plane conductivity on the in-plane TL, with doping dependent
normalization  parameters. This dependence reflects the relationship between the relaxation time,
which determines the in-plane conductivity,  and the phase coherence length.
The data on the electron-doped  $Nd_{2-x}Ce_{x}CuO_{4-y}$ indicate that these crystals are  in
an  intermediate regime in which the c-axis TL is no longer  T-independent but is still
much smaller  and changes with temperature at a smaller rate than the in-plane TL.

%\noindent

%{\footnotesize\it T}

\nonumsection{Acknowledgments}
\noindent
This research was supported by the National Science Foundation under Grant No.
DMR-9801990 at KSU, and the US Department of Energy under  Contract No.
W-31-109-ENG-38 at ANL and Grant No. DE-FG03-86ER-45230 at
UCSD.

\nonumsection{References}

\end{document}